\documentclass[%
 aip,
 amsmath,amssymb,
reprint,%
]{revtex4-1}

\usepackage{graphicx}
\usepackage{dcolumn}
\usepackage{bm}

\usepackage[utf8]{inputenc}
\usepackage[T1]{fontenc}
\usepackage{mathptmx}
\usepackage{etoolbox}

\makeatletter
\def\@email#1#2{%
 \endgroup
 \patchcmd{\titleblock@produce}
  {\frontmatter@RRAPformat}
  {\frontmatter@RRAPformat{\produce@RRAP{*#1\href{mailto:#2}{#2}}}\frontmatter@RRAPformat}
  {}{}
}%
\makeatother

\usepackage{graphicx}
\usepackage[utf8]{inputenc}
\usepackage{amsmath,amssymb,mathtools,bbm}
\usepackage{psfrag}
\usepackage{color}
\usepackage[nice]{nicefrac}
\usepackage[linesnumbered,ruled,lined]{algorithm2e}

\RequirePackage{fix-cm}
\newcommand{\heading}[1]{\medskip\noindent\emph{#1}}

\newcommand{\ddt}[1]{\frac{{\rm d}{#1}}{{\rm d}t}}

\newcommand{\E}{\mathbb E}

\DeclarePairedDelimiterX{\expectarg}[1]{[}{]}{%
  \ifnum\currentgrouptype=16 \else\begingroup\fi
  \activatebar#1
  \ifnum\currentgrouptype=16 \else\endgroup\fi
}
\newcommand{\innermid}{\nonscript\;\delimsize\vert\nonscript\;}
\newcommand{\activatebar}{%
  \begingroup\lccode`\~=`\|
  \lowercase{\endgroup\let~}\innermid 
  \mathcode`|=\string"8000
}

\begin{document}

\preprint{Submit to CHAOS}

\title[Multi-band oscillations emerge from a simple spiking network]{Multi-band oscillations emerge from a simple spiking network}
\author{Tianyi Wu}
 \altaffiliation[]{Equal contribution.}%
 \affiliation{ 
School of Mathematical Sciences, Peking University, Beijing 100871, China;
}%
\affiliation{ 
Center for Quantitative Biology, Peking University, Beijing, 100871, China;
}%
\author{Yuhang Cai}%
\altaffiliation[]{Equal contribution.}%
\affiliation{ 
Department of Mathematics, University of California, Berkeley, CA, USA 94720;
}%
\author{Ruilin Zhang}%
\affiliation{ 
Center for Quantitative Biology, Peking University, Beijing, 100871, China;
}%
\affiliation{ 
Yuanpei College, Peking University, Beijing, 100871, China;
}%

\author{Zhongyi Wang}%
\affiliation{ 
School of Mathematical Sciences, Peking University, Beijing 100871, China;
}%
\affiliation{ 
Center for Quantitative Biology, Peking University, Beijing, 100871, China;
}%
\author{Louis Tao}
\altaffiliation[]{Correponding authors.}%
\affiliation{ 
Center for Quantitative Biology, Peking University, Beijing, 100871, China;
}%
\affiliation{%
Center for Quantitative Biology, Peking University, Beijing, 100871, China;}%
\email{taolt@mail.cbi.pku.edu.cn.}

\author{Zhuo-Cheng Xiao}
\altaffiliation[]{Correponding authors.}%
\affiliation{%
Courant Institute of Mathematical Sciences, New York University, NY, USA 10003.}%
\email{xiao.zc@nyu.edu.}
\date{\today}

\begin{abstract}
In the brain, coherent neuronal activities often appear simultaneously in multiple frequency bands, e.g., as combinations of alpha (8-12 Hz), beta (12.5-30 Hz), gamma (30-120 Hz) oscillations, among others. These population rhythms are believed to underlie information processing and cognitive functions and have been subjected to intense experimental and theoretical scrutiny. Computational modeling has provided a framework for the emergence of network-level oscillatory behavior from the interaction of spiking neurons. However, because of the strong nonlinear interactions between highly recurrent spiking populations, the dynamic interplay between neuronal rhythms in multiple frequency bands has rarely been theoretically investigated. Many studies invoke multiple physiological timescales (e.g., various ion channels or multiple types of inhibitory neurons) or oscillatory inputs to produce rhythms in multi-bands. Here we demonstrate the emergence of multi-band oscillations in a simple network consisting of a single excitatory and a single inhibitory neuronal population driven by constant input. First, we construct a data-driven, Poincaré section theory for robust numerical observations of single-frequency oscillations bifurcating into multiple bands. Then we develop model reductions of the stochastic, nonlinear, high-dimensional neuronal network to capture the appearance of multi-band dynamics and the underlying bifurcations theoretically. Furthermore, when viewed within the dimensionally-reduced state space, our analysis reveals conserved geometrical features of the bifurcations on low-dimensional dynamical manifolds. These results suggest a simple geometric mechanism behind the emergence of multi-band oscillations without appealing to oscillatory inputs or multiple synaptic or neuronal timescales. Thus our work points to unexplored regimes of stochastic competition between excitation and inhibition behind the generation of dynamic, patterned neuronal activities.
\end{abstract}

\maketitle

\begin{quotation}
Coherent neural activities in the brain are well studied because of the belief that they are the dynamical substrate underlying information processing and cognitive functions. Frequently, these neural dynamical patterns appear as synchronous, stochastic, population-level spiking activities. Because of the strong, rapid and recurrent interactions between excitatory and inhibitory neurons, theoretical modeling of these coherent spiking patterns has faced enormous challenges. Here we study a dynamical regime where these synchronous activities appear as collective oscillations in multiple frequencies, even in the presence of temporally constant inputs. Making use of the appearance of nearly periodic dynamical spiking patterns, we combine classical Poincare theory with a data-driven approach to construct a stochastic bifurcation theory to explain the prominence of multi-band oscillations. We develop model reductions to deconstruct the underlying bifurcations. Furthermore, applying our recently developed dimensional reduction methods, these bifurcations can be viewed as conserved geometrical features on low-dimensional manifolds. In particular, we show that multi-band oscillatory activities can be described by a geometry of ``leaves and roots,'' where each ``leaf'' corresponds to a particular   beat, and the ``root'' encapsulates the complex initial conditions that trigger each stochastic synchronous event.
%
%
\end{quotation}

\section{Introduction\label{sec_intro}}
Coherent neuronal activities are found in many brain regions, ranging from visual and auditory cortices \cite{GrayEtAl1989,AzouzGray2000,LogothetisEtAl2001,HenrieShapley2005,BroschEtAl2002}, to frontal and parietal cortices \cite{BuschmanMiller2007,GregorgiouEtAl2009,SiegelEtAl2009,SohalEtAl2009,CanoltyEtAl2010,SigurdssonEtAl2010,van2010learning,PesaranEtAl2002,MedendorpEtAl2007}, to the hippocampus \cite{BraginEtAl1995,CsicsvariEtAl2003,ColginEtAl2009,Colgin2016}, the amygdala \cite{PopescuEtAl2009}, and the striatum \cite{vanderMeerRedish2009}, and are often manifested as neuronal oscillations. Dynamically, these rhythms are likely to be a reflection of the many possible interactions between excitation and inhibition, between disparate synaptic and neuronal timescales, and between local and long-range recurrent connectivities. It is also commonly believed that they form a neurophysiological basis of sensory processing and cognitive functions, such as attention \cite{FriesEtAl2001,FriesEtAl2008}, learning \cite{bauer2007gamma}, and memory \cite{PesaranEtAl2002}. Therefore, neuronal oscillations have been subjected to intense experimental and theoretical examinations, ranging from the specific functions of these brain rhythms to inferences of their roles in information processing and brain computations.

Much experimental evidence has correlated various population oscillations to enhanced cognition or behavioral tasks. For instance, the enhancement of gamma oscillations has been shown to sharpen visual feature detection in V1 \cite{AzouzGray2000,AzouzGray2003,FrienEtAl2000,WomelsdorfEtAl2012} and direction selectivity in MT \cite{LiuNewsome2006,khayat2010frequency}. Attentional states are often accompanied by more significant gamma-band synchronization \cite{FriesEtAl2001,engell2010selective,benchenane2011oscillations,poch2014modulation}. Theta-band oscillations have been found to be correlated with visual spatial memory \cite{kaplan2014medial,herweg2020theta,goyal2020functionally,vivekananda2021theta}, while physiologically and pharmacologically-enhanced theta rhythms improve learning and memory \cite{sweet2014improved,stoiljkovic2015modulation,roberts2018entrainment,herweg2020theta}. Alpha oscillations have been linked to perceptual learning \cite{sigala2014role,poch2014modulation,bays2015alpha,brickwedde2019somatosensory,he2021causal}, and recent work has suggested a role in coordinating activity between brain regions \cite{hindriks2015role,herweg2016theta,zhang2018theta}. Cortical beta rhythms have been correlated with working memory, decision making, and sensorimotor activities \cite{haegens2011beta,kilavik2013ups,tan2016post,schmidt2019beta}.

Many studies, experimental, theoretical, and modeling, have focused on the emergence of collective rhythms of a single frequency band. Numerical simulations of large-scale models of the visual cortex have shown dynamics consistent with the experimentally observed gamma rhythms \cite{chariker2018rhythm}. Studies demonstrated that coherent neuronal oscillations can be driven by external inputs or may emerge as internal states of the network \cite{RubinTerman2004,pmid18480283}. Theoretical investigations have pointed to the importance of recurrent couplings between interneurons and excitatory pyramidal cells  
and the role of synaptic timescales \cite{wang1996gamma,whittington2000inhibition,BorgersKopell2003,chariker2015emergent,bezaire2016interneuronal,chariker2018rhythm,KeeleyEtAl2019}. Substantial theoretical progress has been made by mapping the so-called theta neurons to systems of phase oscillators, which, in the weakly coupled limit, are amenable to mean-field reductions demonstrating how collective dynamics can lead to phase synchronization\cite{AshwinCoombesNicks2016,BickEtAl2020}.

Compared to single-frequency band neuronal rhythms, there has been much less work on the mechanisms underlying multi-band oscillations. Theoretically, studies have built upon single frequency synchronization by introducing multiple synaptic or neuronal timescales or by the interaction between multiple neuronal pools \cite{bezaire2016interneuronal,ter2021comprehensive,zonca2021emergence}. Researchers have also pointed out that different frequency input forcing near the Hopf bifurcation of a Hodgkin-Huxley type neuron can result in nested oscillations, similar to the theta-gamma nested rhythms observed in the hippocampus and during rapid eye movements \cite{jensen2007cross,segneri2020theta}.

Here we demonstrate that multi-band rhythms can emerge from a small neuronal network consisting of a single excitatory population recurrently coupled with a single inhibitory population, both driven by constant input. Recently, we investigated how stochastic gamma oscillations form a low-dimensional manifold in a dimension-reduced state space \cite{CaiWuTaoXiao2021}. Here we show that a more detailed exploration of the dynamics reveals regions in parameter space where single-frequency, stochastic oscillations break up into different frequency bands, even in the presence of temporally stationary inputs.

In the next section, we introduce our leaky integrate-and-fire (LIF) neuronal network model and explore the regime of repetitive, nearly periodic multiple-firing events (MFEs) as a function of model parameters. Focusing first on the repetition, we formulate and examine a data-driven Poincar\'e section theory. 
The Poincar\'e sections allow us to map the iterative, nearly synchronous coherent spiking events in a reduced, low-dimensional space. Our analysis reveals that the multi-band oscillations emerge from different modes of excitation-inhibition (E-I) competitions between neural populations, which are sufficiently captured by membrane potential profiles. 

\section{Results\label{sect_Rslt}}
Experimentally, neuronal oscillations are typically studied by electrophysiological methods, and rhythmic oscillatory behavior is usually observed in the recordings of local field potentials\cite{RayMaunsell2015,HenrieShapley2005}. 
Mathematically, neuronal oscillations were usually modeled by temporally reoccurring activity patterns of a neuronal network. Well-known theories include the interneuronal network gamma (ING \cite{wang1996gamma,Livingstone1996,geisler2005contributions}), the pyramidal interneuronal network gamma (PING \cite{whittington2000inhibition,BorgersKopell2003}), multiple-firing events (MFEs, \cite{chariker2015emergent,chariker2018rhythm}), among others.
Although these models include different levels of biological details and describe various types of oscillations in different brain areas, they generally share two common views: 
First, neuronal oscillations are reflected by reoccurring synchronized spiking patterns generated by neural ensembles;
Second, the reoccurring spiking patterns are induced by the rapid, recurrent interactions between neurons. 
Therefore, the informative aspects of neuronal oscillations (including frequencies, amplitudes, and phases) can be represented by the statistics of the spiking patterns. Furthermore, any study of the mechanism rests on the understanding of the neuronal interactions within the ensemble.

This paper investigates multi-band neuronal oscillations emerging from the dynamics of a simple spiking neuronal network containing only two neuronal populations: excitatory (E) and inhibitory (I), which are recurrently coupled. Oscillatory or any other temporally inhomogeneous inputs are absent. 
Following previous work \cite{ZhangZhouEtAl2014}, we focus on the so-called multiple-firing events (MFEs). We first present a few different temporally repetitive, synchronized spiking patterns found during our parameter space explorations. Appearing with different amplitudes and phases, these reoccurring MFEs form different \textit{beats} robustly in many parts of parameter space and some have as many as three oscillatory bands: alpha (8-12 Hz), beta (12.5-30 Hz), and gamma (30-90 Hz). 

To understand the emergence of different beats, we need to explain how the repetitions and alternations of MFEs are shaped by the tight competition between E/I populations. 
Here, we develop a novel data-driven Poincaré section theory for the network dynamics.
Most remarkably, our Poincaré section reveals that the E-I competition is essentially captured by the membrane potential profiles just before the occurrence of each MFE. 
Furthermore, by considering only the most informative features of the membrane potential distributions, a stochastic bifurcation analysis illustrates that 
\begin{enumerate}
    \item Different beats are shaped by different balances of the E-I competition arising from the bifurcations;
    \item The multi-band oscillations are prevalent in the high-dimensional parameter space.
\end{enumerate}
In addition, when presented in an appropriate dimensionally-reduced state space, these MFE beats collectively form notable geometrical structures on the principal manifolds. This novel representation of spiking network dynamics reveals global features emerging from local bifurcations.

The above physics \& data-informed model reduction provides a \textit{top-down} view of the emergence of multi-band oscillations from the detailed LIF model. 
To the opposite, we also develop an autonomous ODE model from a \textit{bottom-up} perspective with abbreviations of most of the biophysical details, only preserving the most crucial features for oscillations. Notably, it qualitatively reproduces the MFE beat patterns and bifurcations by using only the low-order moments of the membrane potential distributions.
In all, the model reduction and ODE model meet in the half way and exhibit surprisingly similar geometrical features in the reduced phase space, confirming the tight relation between multi-band oscillations and delicated competitions between E and I populations. 

Taken together, these results provide a theoretical framework to analyze the complicated oscillatory dynamics produced by the rapid competition between the excitatory and inhibitory neuronal populations.
\begin{figure*}
  \begin{center}
    \includegraphics[width=0.9\textwidth]{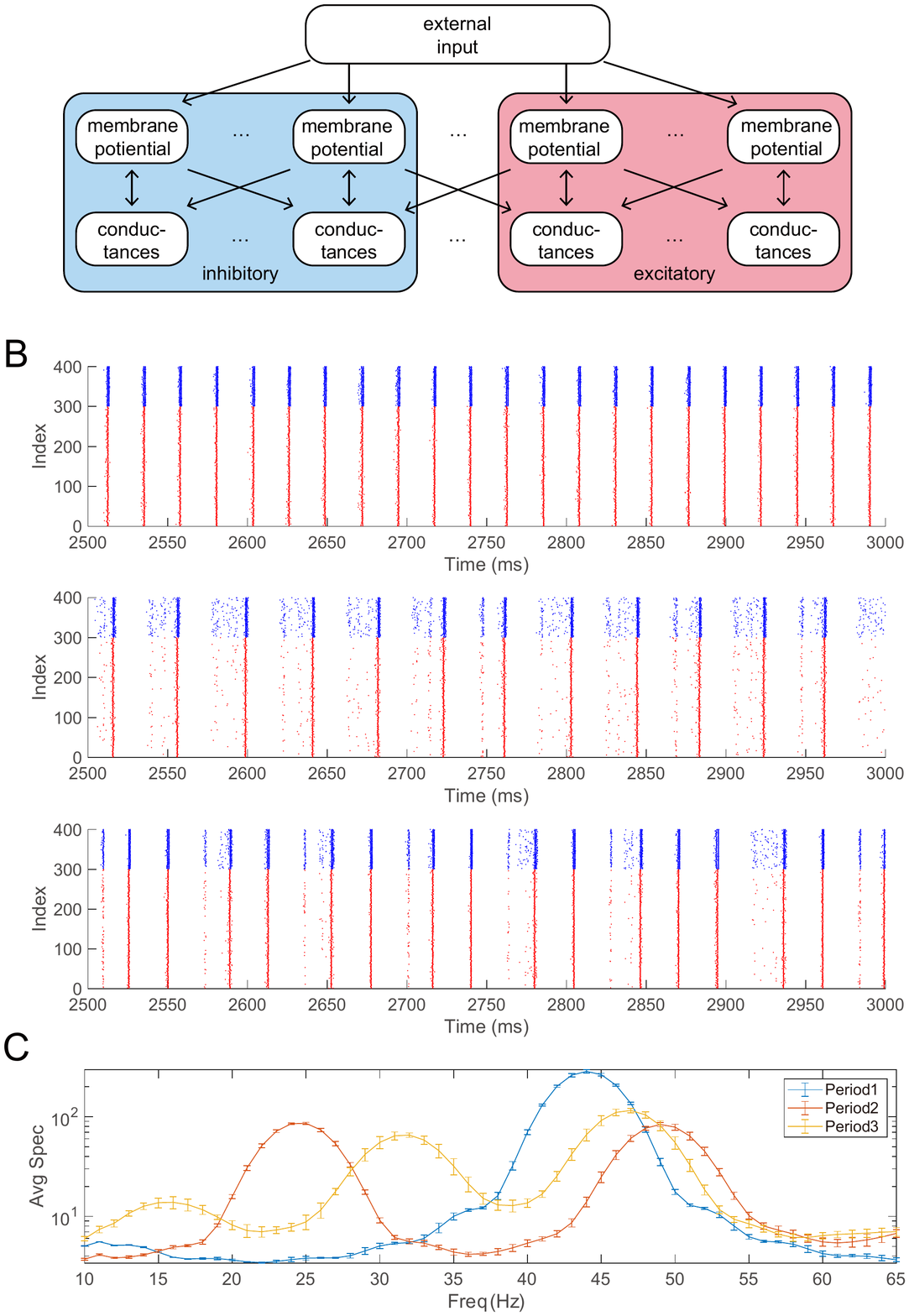}
    \caption{A network model producing different oscillatory dynamics.  \textbf{A.} Network structure of the Leaky-Integrate-Fire network.  \textbf{B.} Raster plots exhibiting three different classical temporal repetitive patterns of MFEs: periods 1, 2, and 3. Excitatory/inhibitory spikes are indicated by red/blue dots. \textbf{C.} Temporally averaged frequency spectrums of the three MFE patterns in \textbf{B}. The error bars indicate the standard error. }
    \label{Fig1: Model} 
  \end{center}
\end{figure*}
\subsection{Synchronized firing in a small integrate-and-fire network}
We numerically simulate a small, 400-neuron network that has been shown to produce stochastic, oscillatory dynamics in the gamma band \cite{chariker2015emergent,li2019well,li2019stochastic,CaiWuTaoXiao2021}. The parameter values were taken from our previous studies \cite{CaiWuTaoXiao2021} of an input layer (Layer 4C$\alpha$) of the macaque primary visual cortex (V1). We will demonstrate that significantly different temporal repetitions of spike patterns can be produced by varying the neuronal and network parameters systematically near the parameter values that well-model many V1 electrophysiological data. 

\subsubsection{A simple network model}
The network contains two populations of spiking neurons: 300 excitatory (E) cells and 100 inhibitory (I) cells. The network structure is assumed to be spatially homogeneous, while the synaptic couplings and single neuron dynamics are intentionally fixed near previous models \cite{li2019well,li2019stochastic,CaiWuTaoXiao2021}. 
All neurons are driven by independent, identically distributed Poisson spike trains. Here the strength of this feedforward input maintains each neuron at a physiologically realistic firing rate.

The network dynamics is determined by the membrane potential ($v$) of each neuron, which is modeled by a linear, leaky integrate-and-fire (LIF) equation:
\begin{equation}
    \label{Rslt:Eq_LIF}
    \ddt{v} = -g^{L}(v-V_r)-g^E(v-V^E)-g^I(v-V^I). 
\end{equation}
The membrane potential $v$ is a dimensionless variable ranging between the reversal potential $V^I = -2/3$ and the spiking threshold $V_{th} = 1$. $v$ is driven towards $V_{th}$ by the excitatory current $g^E(v-V^E)$, and away from it through the leaky and inhibitory currents ($g^{L}(v-V_r)$ and $g^I(v-V^I)$). In LIF dynamics, when the membrane potential $v$ arrives at $V_{th}$, a spike is emitted by this neuron to all its postsynaptic cells, concurrently $v$ is reset to the rest potential $V_r = 0$, and then held there for an absolute refractory period $\tau^{\textrm{ref}}$. The E and I conductance terms are sums of Green's functions of the spiking series received by this neuron. The key features of the computation of the LIF neurons are depicted in Fig.~\ref{Fig1: Model}A and detailed in \textbf{Methods}. 

\subsubsection{Different MFE beats}
\label{Rslt_Sect1_1}
The network described above has been shown to produce stochastic gamma-band oscillations in many previous studies \cite{li2019well,li2019stochastic,CaiWuTaoXiao2021}. Namely, spikes of E and I neurons cluster in time as MFEs, which reoccur within timescales of roughly $\sim$40 Hz (see, e.g., the Fig.~\ref{Fig1: Model}B, upper panel). Despite the stochastic external drive, previous work demonstrated that the robust recurrence of MFEs is due to the synaptic timescales during the dynamical competition between the E and I populations \cite{chariker2018rhythm,KeeleyEtAl2019,li2019well,CaiWuTaoXiao2021}. Specifically, an MFE is triggered by the stochastic firing of a couple of E cells, leading to the rapid recruitment of both E and I neuron firings. However, the I-current peaks later than the E-current because of the longer timescale of the inhibitory synapse (see \textbf{Methods}). Therefore, in typical network dynamics involving gamma oscillations, both E and I populations initially undergo highly correlated firing, then followed by a dominance of inhibition, resulting in the termination of the MFE. After the inhibition wears off, the network again enters an excitable state, allowing for the next occurrence of MFE. We refer the readers to \cite{chariker2015emergent} for more details of this recurrent excitation-inhibition mechanism of stochastic gamma oscillations. 

However, the recurrence of MFE does not necessarily result in consecutively similar spiking patterns. Instead, we find that MFEs may recur in different beats and rhythms.
During our parameter space investigation, the amplitudes of MFEs (i.e., number of neurons involved) repetitively alter in different fashions.
We describe the three most prevalent repetitive patterns of MFEs below.

The first pattern depicts nearly regular recurrence of MFEs in the gamma band, where consecutive MFE amplitudes do not vary significantly (Fig.~\ref{Fig1: Model}B, upper). 
This classical 1-beat repetitive pattern has been well studied in previous literature associated with gamma oscillations \cite{chariker2015emergent,li2019well,li2019stochastic,CaiWuTaoXiao2021}. 
To the opposite, consecutive MFEs can alternate with large and small amplitudes and form multiple beats. 
For example, in a 2-beat rhythm, a strong MFE is usually followed by a weak one and vice versa (Fig.~\ref{Fig1: Model}B, middle). On the other hand, a weak MFE may occurs after every two consecutive strong MFEs (Fig.~\ref{Fig1: Model}B, bottom), forming a 3-beat repetitive pattern.
We remark that the three distinct MFE rhythms can take place when varying only one physiological parameter (in Fig.~\ref{Fig1: Model}B, the cortical coupling I-to-E weights $S^{EI}$: 1-beat, $S^{EI} = 2.01$; 3-beat, $S^{EI} = 2.07$; 2-beat, $S^{EI} = 2.16$).

These repetitive patterns appear as different numbers of peaks in the temporal frequency spectrum (Fig.~\ref{Fig1: Model}C). Compared to the single peak $\sim$45 Hz for the 1-beat rhythm, the alternation of MFE amplitudes in the 2-beat rhythm introduces an additional peak $\sim$25 Hz (high beta band), and the 3-beat rhythm results in one more $\sim$15 Hz (low beta band). At the same time, the gamma-band peak is shifted towards higher frequencies as the network recovers more rapidly from the weaker inhibition induced by weaker MFEs.

Needless to say, the rhythms listed above was only a small subset of all possible network dynamics. On the other hand, the patterns were the most prominent ones during our parameter space exploration (via 1-dimensional sweeps of a few critical parameters; see Sect \ref{Sect:Rslt-bifurcation}). Thus, in the remainder of this paper, we focus on these highly robust and persistent MFE repetitive patterns and leave more thorough investigations of the high-dimensional parameter space to future work. 

\subsection{Multi-bands arise from iterations of MFEs}
The regularity of MFE recurrence as different beats suggests the importance of dynamical paths between consecutive MFEs.
Specifically, how is the next MFE induced by the E-I competition determined by the current MFE? 
We demonstrate that the dynamical paths can be captured via an iterative scheme. 
The idea is intuitively simple: If one could predict the next MFE given all the information of the current MFE, say, from a mapping $F$, then the long-term network dynamics are revealed by $F^n$, i.e., all subsequent MFEs are predicted iteratively.

To address the mechanism behind different MFE beats, this section proposes a data-driven Poincaré section theory regarding MFE iterations. 
We will perform a top-down analysis: Starting from the high-dimensional network dynamics, we deduce and simplify the MFE mapping $F$ by combining first principles arguments and data-informed numerical simulations.

\begin{figure*}
  \begin{center}
    \includegraphics[width=0.9\textwidth]{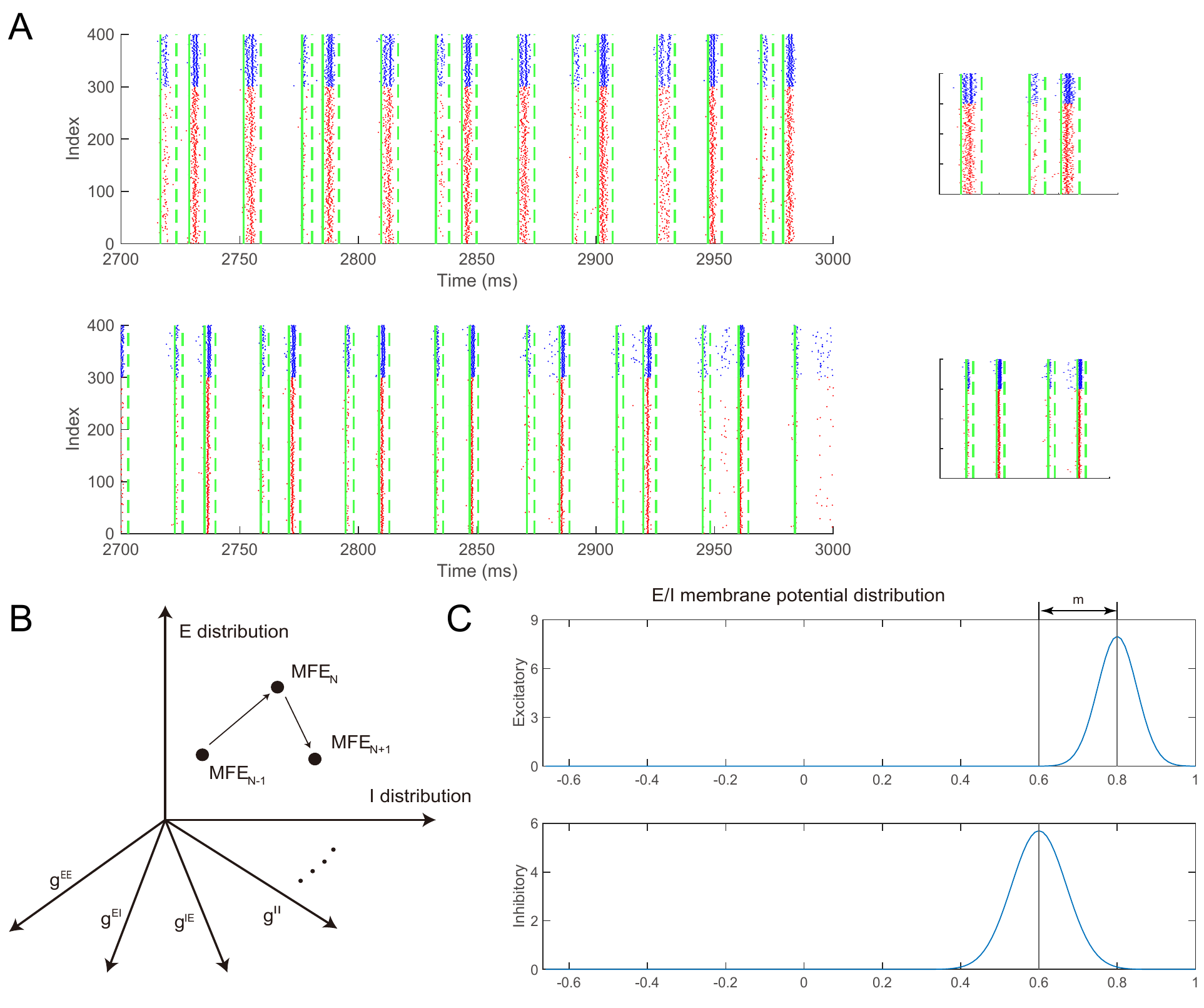}
    \caption{A theoretical framework to analyze the repetitive patterns of MFEs.  
    \textbf{A.} The raster plots are divided into MFEs (the accumulated dots) and inter-MFE intervals (relatively quiescence periods) by a MFE-detection algorithm. The initiation/termination of an MFE is labeled by a solid/dash green line. 
    Two different repetitive patterns are depicted in this panel to demonstrate the effectiveness of the algorithm. 
    \textbf{B.} The iteration from one MFE to another equivalents to a stochastic, high-dimensional discrete dynamical system. 
    \textbf{C.} The two E/I population potential profile distributions assumed to be truncated Gaussian before the initiation of MFEs. The distribution with a higher average has a support whose upper bound intersects the firing threshold ($V_{th}=1$). The difference between the averages is indicated ($m$).} 
    \label{Fig2: MFE_Iter} 
  \end{center}
\end{figure*}

\subsubsection{Reduction of a high-dimensional MFE mapping}
\label{Sect-3.2.1-MFE_Mapping}
We start by dissecting the network dynamics onto a sequence of time intervals $\{I_n = [t_n, t_{n+1})\}$, where $t_n$ is the starting time of the $n$-th MFE (see definition in \textbf{Methods}). 
Without justification, we first assume that there exists an MFE mapping $F$ capturing the dynamical paths from one MFE to another. Thus, the first natural question is:
How $F$ predicts the dynamical path on $I_{n+1}$ given the network dynamics on interval $I_n$?

Despite its complexity, the dynamical path on $I_n$ is determined by two components: 
\begin{itemize}
    \item The network state on time section $t_n$ (denoted as $\mathbb{M}_n$). $\mathbb{M}_n$ is the initial condition of the dynamical path. 
    \item The realization of all stochasticity $\xi$ within interval $I_n$, including noises in external drives and internal randomness in synaptic couplings.
\end{itemize}
Consequently, the mapping $F$ between dynamical paths is equivalent to a stochastic Poincaré section mapping $F_1$ concerning the network states  
\begin{equation}
    \label{Rslt:Eq_MFE}
    \mathbb{M}_{n+1} = F_1(\mathbb{M}_n, \xi),
\end{equation}

The network state $\mathbb{M}_n$ includes membrane potentials and E/I synaptic conductances $(v, g^E, g^I)$ of all neurons.
Even for our small 400-neuron network, $\mathbb{M}_n$ contains at least $3\times 400 = 1200$ variables. Therefore, it would be hard to study $F_1$ without a drastic dimension reduction.
To make it worse, MFEs are produced by rapid, transient competitions between E/I populations \cite{chariker2015emergent,li2019well,li2019stochastic,CaiWuTaoXiao2021}. Therefore, each MFE can be very sensitive to random events as small as a couple of synaptic failures.
In all, a first-principles-based analytic description of MFE mappings is unfeasible to the best of our knowledge. 

Instead, we propose to investigate $F_1$ in a \textit{data-informed} manner:
We first demonstrate a two-level model reduction of the original mapping $F_1$, then reconstruct the reduced version of $F_1$ from surrogate data.

\heading{Step 1: Coarse-Graining.}
Due to the homogeneous network structure, we coarse-grain the spiking network by assuming the interchangeability between neurons with similar potential profiles.
Instead of concerning the triplet $(v, g^E, g^I)$ for every neuron, the coarse-grained network state is completely determined by the population statistics of the membrane potentials ($\rho^E(v)$, $\rho^I(v)$) and the total conductances $g^{EE}$, $g^{EI}$, $g^{IE}$, $g^{II}$. Here, $g^{PQ}$ represents the sum of $Q$-conductances of all type-$P$ neurons ($P,Q\in{E,I}$).
Therefore, the mapping $F_1$ is simplified to
\begin{equation}
    \label{Rslt:Eq_MFE_CG}
    M_{n+1} = F_2(M_n, \xi),
\end{equation}
where $M_n = (\rho^E(v),\rho^I(v),g^{EE},g^{EI},g^{IE},g^{II})$ at time section $t_n$. 
Fig.~\ref{Fig2: MFE_Iter}B illustrates the schematics of the iteration in this coarse-grained state space.
We remark that, through Eq.~\ref{Rslt:Eq_MFE} and \ref{Rslt:Eq_MFE_CG}, the E-I competition during MFEs are effectively represented by the distributions of E/I membrane potential profiles.

\heading{Step 2: Gaussianity of $\rho^{E,I}(v)$.} The coarse-grained network state $M_n$ still includes four variables (conductances) and two distributions (of membrane potential profiles). To reduce further the complexity of MFE mappings, we impose three additional assumptions to network states on time section $t_n$.
\begin{enumerate}
    \item $g^{PQ}\approx 0$ at the beginning of an MFE ($P,Q\in\{E,I\}$) since the durations of inter-MFE intervals are generally 3-6 times of the longest synaptic timescale (see details in \textbf{Methods}); 
    \item The membrane potential distributions $\rho^E(v)$ and $\rho^I(v)$ are well approximated by truncated Gaussian distributions whose variances are dominated by the noise of external input (Fig.~\ref{Fig2: MFE_Iter}C);
    \item There is at least one neuron whose membrane potential is near the firing threshold $V_{\textrm{th}}$.
\end{enumerate}

Assumptions 1\&2 come from the lack of recurrent drives during the inter-MFE intervals (Fig.~\ref{Fig2: MFE_Iter}A). As observed in the network simulations, the number of spikes produced outside MFEs is small compared to spikes during MFEs, and therefore, the effects of the network spikes produced during inter-MFE intervals on potential distributions and conductances can be neglected.
Assumption 3 comes from the definition of $t_n$: it would take more time after $t_n$ to trigger the first spike of MFE if no neuron had a potential profile at $V_{\textrm{th}}$.
According to the truncated Gaussian assumption (assumption 2), at $t_n$, the upper bound of at least one \textit{support} of $\rho^{E,I}(v)$ should be $V_{\textrm{th}}$ (Fig.~\ref{Fig2: MFE_Iter}C). 

Therefore, for the coarse-grained network state $M_n$, the information provided by $\rho^{E,I}(v)$ equivalents to the positions of the two truncated Gaussians.
In other words, the only relevant statistics is the difference between the averages of the two membrane potentials profiles: $m_n = \E[v^E(t_n)] - \E[v^I(t_n)]$.
In all, $F_2$ can be further reduced to a 1-dimensional mapping $f$, such that
\begin{equation}
    \label{Rslt:Eq_Mean_iter}
    m_{n+1} = f(m_n, \xi),
\end{equation}

Eq.~\ref{Rslt:Eq_Mean_iter} is a drastic simplification compared to Eq.~\ref{Rslt:Eq_MFE_CG}.
We hope $m_n$ alone can capture sufficient information to reproduce the observed MFE beats, whose dynamics are dictated by the strong E-I competition during and after MFEs.

\heading{Step 3: Reconstruct the reduced MFE mapping.}
We now turn to reconstruct $f$ from surrogate data. (We denote the reconstructed version as $\hat{f}$.) 
For any $m_0$ in the relevant domain, we perform a series of Monte Carlo simulations of the transient network dynamics between the initiations of two MFEs. Each simulation contains one independent realization of random paths $\xi$ on $I_0 = [t_0, t_1)$.

Specifically, for each Monte Carlo simulation, we generate membrane potentials for each E/I neuron as the initial condition, then simulate the network dynamics until the next MFE is detected. 
Since it is important to ensure that an MFE is triggered at $t_0$, we sample membrane potentials randomly from two truncated Gaussian distributions $\rho^E(v,t_0)$ and $\rho^I(v,t_0)$ respectively, where $m_0 = \E[v^E(t_0)] - \E[v^I(t_0)]$, and the maximum of membrane potentials is fixed as $V_{\textrm{th}}$.
At the start of the second MFE ($t_1$), we collect the empirical distributions of E/I population potential profiles ($\hat{\rho}^E(v,t_{1})$ and $\hat{\rho}^I(v,t_{1})$).
Finally, the output of $\hat{f}$ from the Monte Carlo simulation is
\begin{align}
    \label{Rslt:Eq_Mean_MC}
    \hat{m}_{1} = \E[\hat{v}^E(t_{1})] - \E[\hat{v}^I(t_{1})],
\end{align} 
where $\hat{v}^{E,I}\sim\hat{\rho}^{E,I}(v,t_{1})$. 

Taking together the outcomes of all Monte Carlo simulations, the output of the stochastic MFE mapping $\hat{f}$ is an empirical distribution of $\hat{m}_{1}$. For more details, see \textbf{Methods}.

Here we must point out that, without the reduction of \textit{Step 2}, a much larger ensemble of surrogate data would be necessary to produce an effective, low-dimensional representation of the MFE mapping. Specifically, if the construction goal was $F_2$, the corresponding Monte Carlo simulations should cover the high-dimensional state space of $M_0$ to estimate the two potential profile distributions and four total conductances. The surrogate data set would be unfeasibly large.

\begin{figure*}
  \begin{center}
    \includegraphics[width=0.9\textwidth]{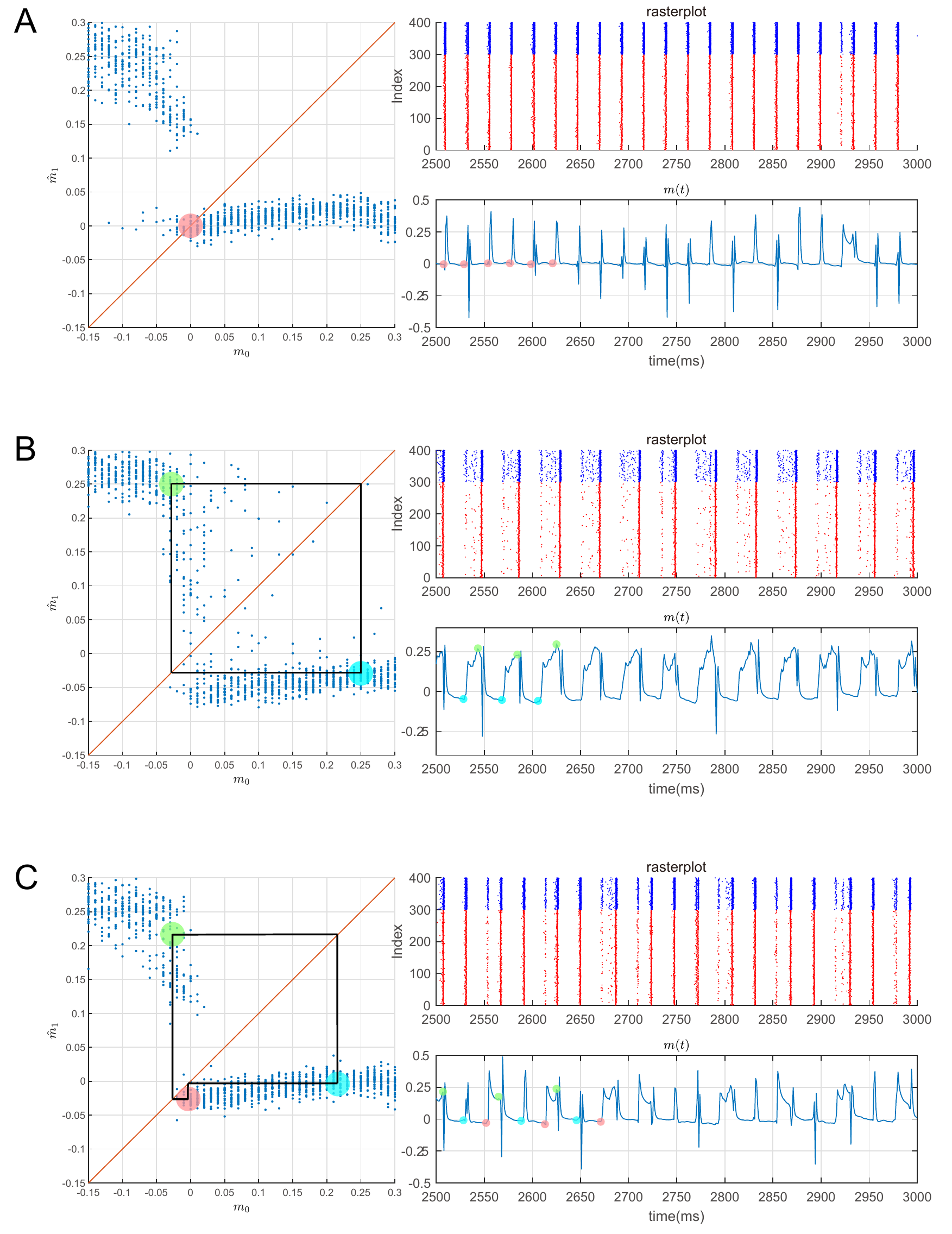}
    \caption{Different repetitive patterns are explained by reconstructed MFE mapping $\hat{f}(m_n, \xi_n)$.  
    Panels \textbf{A-C} correspond to the three repetitive patterns illustrated in Fig.~\ref{Fig1: Model}B, generated by different choices of $S_{EI}$. 
    \textbf{A.} $S_{EI} = 2.01$. 
    Left: Stochastic mapping $\hat{f}(m_n, \xi_n)$ represented by Monte Carlo simulations (blue dots).  Line $x = y$ (red diagonal line) and the only subgroup of the invariant set $\chi_{\hat{f}}$ (pink circle) also indicated. 
    Right Up: Raster plot exhibiting MFEs. 
    Right Bottom: Time trace of $m(t) = \E[v^I(t)] - \E[v^E(t)]$.  The first few $m_n$s are indicated by pink circles at the initiation of each MFE.
    \textbf{B.} $S_{EI} = 2.16$.
    Same as \textbf{A}, but the $\chi_{\hat{f}}$ includes two subgroups (green and pink circles).
    \textbf{C.} $S_{EI} = 2.07$.
    $\chi_{\hat{f}}$ includes three points (green, cyan, and pink circles) and indicates the 3-beat rhythm repetitive pattern.}
    \label{Fig3: Iters}
  \end{center}
\end{figure*}

\vspace{0.5cm}
We now reconstruct $\hat{f}$ aiming for the three MFE repetitive patterns depicted in Fig.~\ref{Fig1: Model}B. 
For each specific $m_n$, 24 Monte Carlo simulations are performed, and the corresponding $\hat{m}_{n+1}$ values are indicated by blue dots in the left plots in Fig.~\ref{Fig3: Iters}A-C. 
We find that the invariant sets of mapping $\hat{f}$ (denoted $\chi_{\hat{f}}$) are clustered as 1, 2 or 3 subgroups (pink, green, and cyan circles, Fig.~\ref{Fig3: Iters}A-C left panels). 
Remarkably, the number of subgroups predicts the beat number of MFE repetitive patterns (Fig.~\ref{Fig3: Iters}A-C, upper right raster plots).

Moreover, the invariant sets $\chi_{\hat{f}}$ agree in detail with the network dynamics.  
More precisely, for each parameter set, $\chi_{\hat{f}}$ fits the membrane potentials E/I population at the initiation of each MFE. (Fig.~\ref{Fig3: Iters} A-C, lower right panels. Colored circles correspond to subgroups of $\chi_{\hat{f}}$ indicated by the same colors).
These results confirm that, in the selected parameter regimes, the 1-dim mapping $\hat{f}$ successfully captures the E-I competition and the underlying mechanism of MFE beats by only considering the averages of potential profiles. 

Furthermore, different MFE beats are well predicted by the geometrical features of $\hat{f}$. 
We note that the lower branch of $\hat{f}$ (when $m_n>0$) undergoes downward spatial translation when $S^{EI}$ increases (Fig.~\ref{Fig3: Iters} left panels. $S^{EI}$ values: $A<C<B$). 
Namely, when $m_n$ is fixed, a stronger $S^{EI}$ coupling means inhibition produced by the MFEs has a more significant impact on the E neurons. 
Therefore, the average of E-population potential profiles $\E[\hat{v}^E]$ will be lower when the next MFE is initiated, leading to a lower $m_{n+1}$.
Collectively, the downward translation of $\hat{f}$ explains the different MFE repetitive patterns: 
From A to C, the only cluster (pink) loses stability and introduces two other clusters (cyan and green) into $\chi_{\hat{f}}$ as its predecessor and successor in the MFE iteration, giving rise to 3-beat in the oscillation;
From C to B, the further lowered branch exhibits no intersections with the diagonal $m_{n+1} = m_n$ (red line), and the remaining two clusters form the 2-beat rhythm oscillation.

\subsubsection{Multi-band oscillations caused by bifurcations}
\label{Sect:Rslt-bifurcation}
\begin{figure*}
  \begin{center}
    \includegraphics[width=\textwidth]{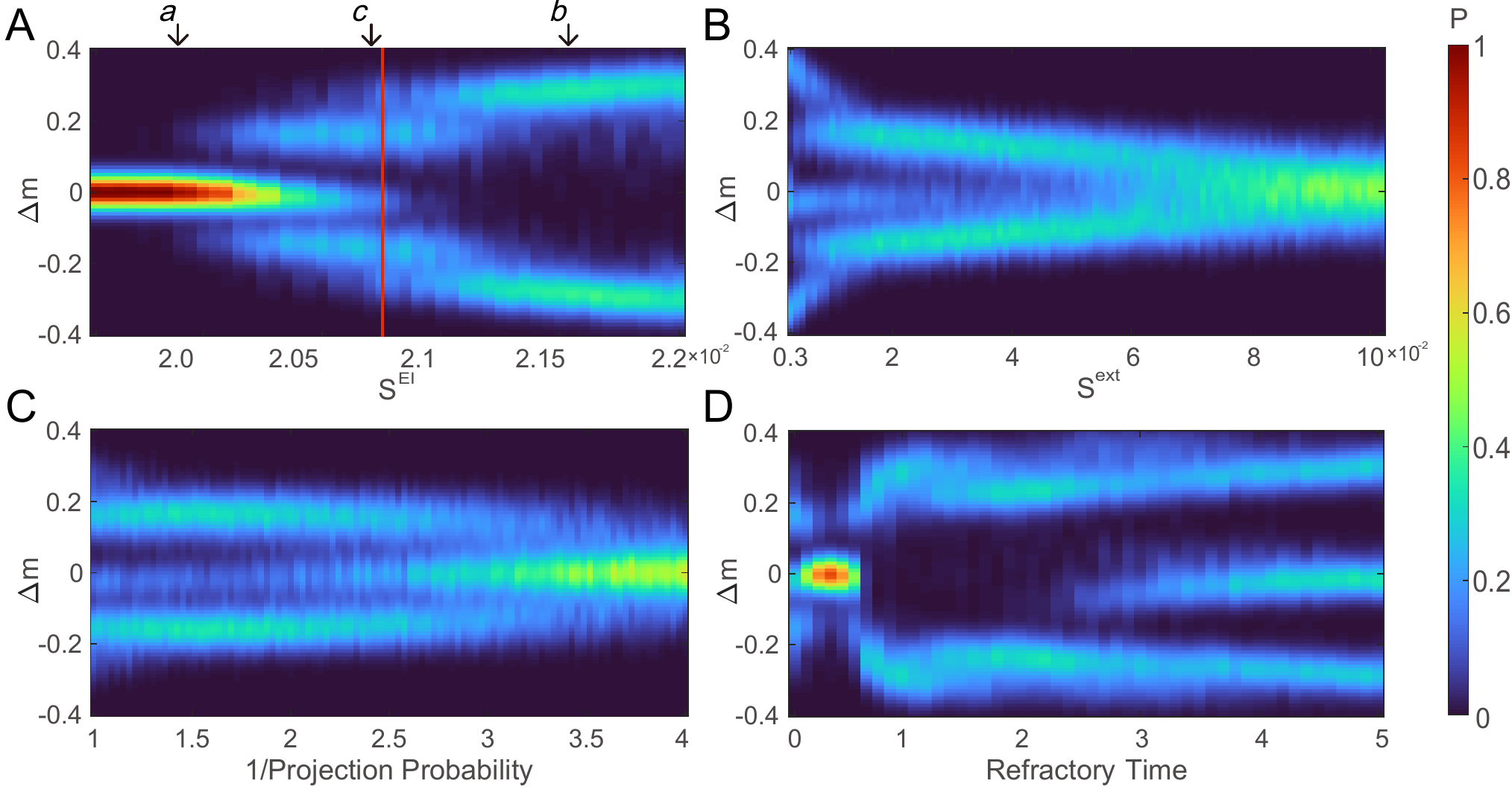}
    \caption{MFE beat bifurcation maps indicated by the distribution of $\rm \Delta m$. 
    \textbf{A}: Bifurcation map of $S^{{EI}}$. $S^{{EI}}$ values of the three patterns in Fig.~\ref{Fig3: Iters} are indicated by arrows.
    \textbf{B-D}: Bifurcation maps of $S^{\rm{ext}}$, $1/p$ and $\tau^{\rm{ref}}$, where $S^{{EI}}$ is fixed as indicated by the red line in panel A. }
    \label{Fig4: Probability density map}
  \end{center}
\end{figure*}

The connection between $\hat{f}$ and MFE beats suggests that multi-band neural oscillations are induced when the MFE mapping undergoes critical bifurcations in the parameter space. 
Here, we argue and examine this point using the reconstructed $\hat{f}$ to investigate 1D sections of the parameter landscapes. 
This subsection aims to demonstrate the prominence of multi-band oscillations in the parameter space of the simple network. 

Here, we focus on a handful of parameters selected from the following groups, since a thorough investigation of high-dimensional parameter landscape is computationally unfeasible. 
\begin{enumerate}
    \item Cortical synaptic couplings. We choose $S^{EI}$, the I-to-E synaptic strength.
    \item External drive. We choose $S^{\textrm{ext}}$, the strength of each external kick. The mean of external stimuli $S^{\textrm{ext}}\cdot\lambda^{\rm{ext}}$ is fixed during the variation of strength.
    \item Network architecture. In a homogeneous network, this is specified by $p$, he connection probability $P$ between each pair of neurons. Likely, $P\cdot S^{Q'Q}$ is fixed for $Q,Q'\in\{E,I\}$ to control the total postsynaptic current induced by each spike.
    \item Neuronal parameters that govern single-cell dynamics. We choose $\tau^{\textrm{ref}}$, the refractory period of a cell.
\end{enumerate}

The four parameters listed above $\{S^{EI},S^{\textrm{ext}},p,\tau^{\textrm{ref}}\}$ are varied individually to investigate the 1-dimensional bifurcations of $\hat{f}$ . 
For each parameter set involved, we reconstruct $\hat{f}$ and expect the invariant set $\chi_{\hat{f}}$ to reflect the MFE beats shown in Sect~\ref{Sect-3.2.1-MFE_Mapping}.
In addition, we propose a \textit{differential visualization} of $\chi_{\hat{f}}$ by concerning $$\Delta m_n = m_{n+1} - m_{n} = (\hat{f}-\mathbbm{1})(m_{n})$$ instead of $m_n$, where $\mathbbm{1}$ stands for the identity function.
The reason is that some subgroups of $\chi_{\hat{f}}$ may occupy similar locations when displayed on the 1D space of $m_n$.
In the differential visualization, we use the subgroup number of $\chi_{\hat{f}-\mathbbm{1}}$ to represent the subgroup number of $\chi_{\hat{f}}$ (Fig.~\ref{Fig4: Probability density map}). 
In all, the MFE beats bifurcations are depicted by the heat maps of empirical distributions of $\chi_{\hat{f}-\mathbbm{1}}$. 

Fig.~\ref{Fig4: Probability density map}A illustrates a systematic investigation of MFE beats bifurcation for different $S^{EI}$, while all other parameters are fixed (same as Fig.~\ref{Fig3: Iters}). 
This confirms our previous finding of the ``1-3-2" beat bifurcation when $S^{EI}$ increases: 
$\chi_{\hat{f}-\mathbbm{1}}$ exhibits only one subgroup for $S^{EI}<2.0$, two subgroups for $S^{EI}>2.1$, and three subgroups in the transition period between them. 
Notably, the unstable subgroup of the 3-beat rhythm (pink, Fig.~\ref{Fig3: Iters}C) is affected by the stochasticity nature of $\hat{f}$. 
Therefore, instead of standing alone, the 3-beat rhythms is always mixed with 1-beat or 2-beat rhythms (see e.g., Fig. \ref{Fig3: Iters}C raster plot, 2650-2800 ms).
This leads to an uneven distribution of mass within the three subgroups of $\chi_{\hat{f}-\mathbbm{1}}$.
For example, when $S^{EI}\approx2.0$, though two other subgroups already emerge (blue and green circles, Fig.~\ref{Fig3: Iters}C), the subgroup around $(0,0)$ still carries most of the mass (pink circle), i.e.,  
$\Delta m_n = m_{n+1} - m_{n}$ is mostly 0 for any consecutive two MFEs.
As a result, the network dynamics are dominated by 1-beat rhythms, with 3-beat rhythms taking place occasionally.
On the opposite, when $S^{EI}\approx2.1$, as indicated by the branches of $\Delta m_n$, the subgroup around $(0,0)$ almost vanishes, and the network dynamics is dominated by 2-beat rhythm patterns instead.

The bifurcation map of $S^{EI}$ suggests that 1- and 2-beat rhythms are more robust, whereas the 3-beat is a more transient phenomenon. 
Therefore, we next focus on how other parameters may affect the bifurcations for 3-beat rhythms.
Three 1-dim parameter scan are carried out ($S^{\textrm{ext}}$, $p$, and $\tau^{\textrm{ref}}$) by fixing $S^{EI} = 2.08$ where 3-beat rhythm patterns are prominent. 

We first examine the bifurcations induced by the noise of external and internal stimulus ($S^{\textrm{ext}}$ and $p$): 
The strength of external stimuli $S^{\textrm{ext}}$ represents the size of external noise, whereas projection probability between neurons $p$ measures the internal variability of network dynamics.
Remarkably, the network dynamics are dominated by 1-beat when external and internal noises are large (high $S^{\textrm{ext}}$ and $1/p$), and dominated by 3-beat rhythm patterns when noises are small (low $S^{\textrm{ext}}$ and $1/p$, Fig.~\ref{Fig4: Probability density map}BC).
These beat bifurcations cohere with our intuition that 3-beat rhythm are disrupted by larger randomness. 

The bifurcations induced by $\tau^{\textrm{ref}}$ are much more intriguing (Fig.~\ref{Fig4: Probability density map}D). 
While dominated by 3-beat rhythm patterns when $\tau^{\textrm{ref}} \sim 0$, the cluster $\Delta m \approx 0$ carries most mass of $\chi_{\hat{f}-\mathbbm{1}}$ for $0.2<\tau^{\textrm{ref}} < 0.8$ ms.
Hence MFEs reoccur in 1-beat rhythms due to $m_{n+1} \approx m_n$.
When the $\tau^{\textrm{ref}}$ becomes longer, the dynamics is dominated by 2-beat rhythms for $0.8< \tau^{\textrm{ref}} < 2.5$.
More interestingly, 3-beat rhythms come back robustly when the refractory becomes sufficiently long.
Though it is so far difficult to provide even a qualitative explanation of this sophisticated bifurcation phenomena, we point out that the bifurcation points of $\tau^{\textrm{ref}}$ locate at the same magnitudes of the synaptic timescales (1-10 ms, see \cite{koch2004biophysics}).
This intuition suggests the refractory periods plays an essential role in the interplay of E-I competition, leading to different MFE beats.
In all, Fig.~\ref{Fig4: Probability density map}D illustrates a crucial dependence of oscillation features on the interplay between physiological timescales, as we have investigated in our previous study \cite{CaiWuTaoXiao2021}. 

\subsubsection{Geometrical features of the bifurcations}

\begin{figure*}
  \begin{center}
    \includegraphics[width=\textwidth]{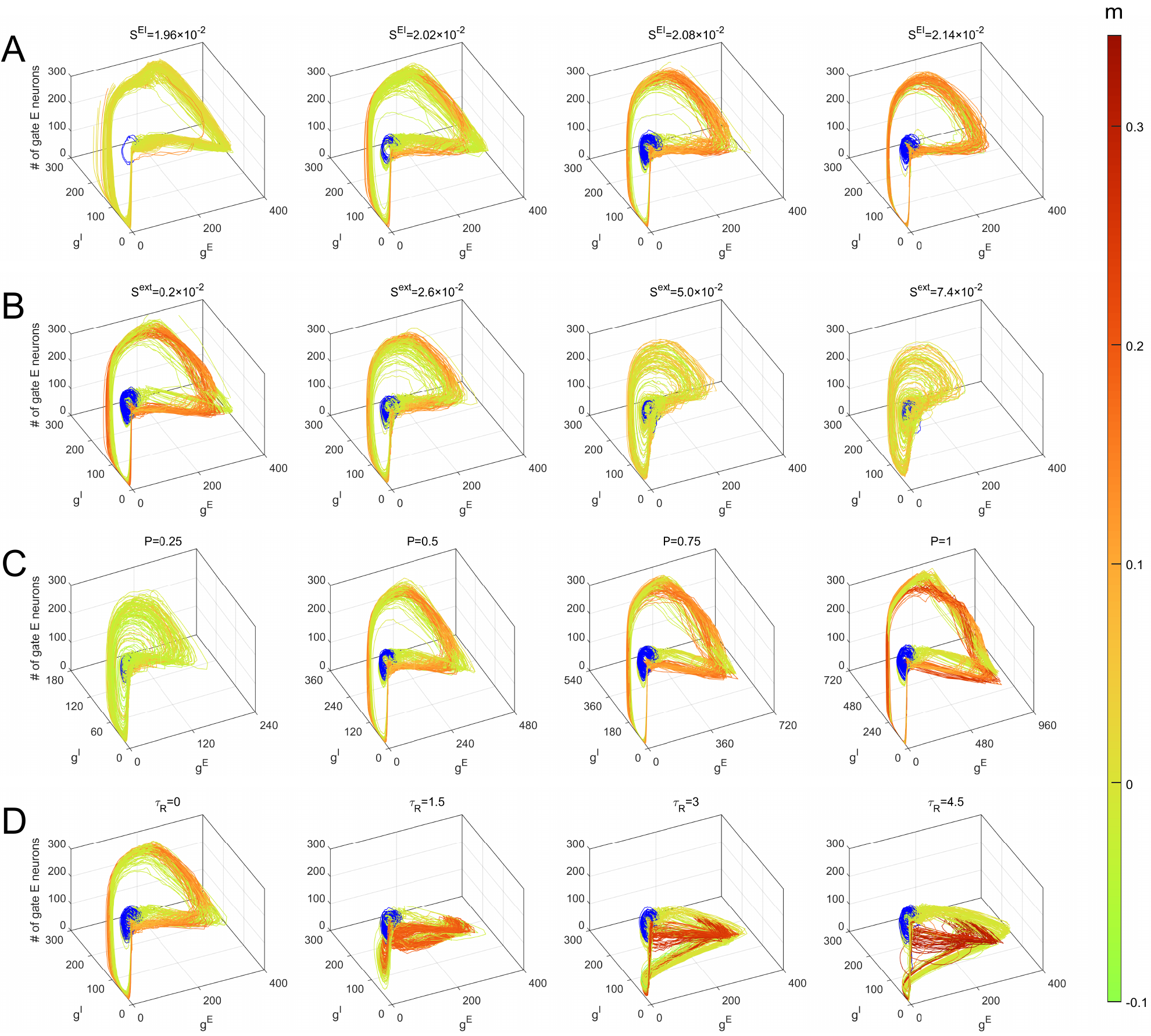}
    \caption{Multiband oscillations produced by network simulation are presented in a reduced 3D space classified by the coarse-grained model in \cite{CaiWuTaoXiao2021}. Trajectories of different beats are colored according to values $\Delta m$ at the initiations of each MFE. Weak beats are specially labeled with deep blue color. \textbf{A.} Four points in the 1D parameter space of $S^{EI}$ are selected. The corresponding network dynamics form a "root-leaf" structure, and the number of leaf branches indicates the MFE beat number demonstrated by the bifurcation map in Fig.~\ref{Fig4: Probability density map}A. \textbf{B-D.} Similar as \textbf{A}, but the varying parameters correspond to Fig.~\ref{Fig4: Probability density map}B-D.} 
    \label{Fig5: Manifold}
  \end{center}
\end{figure*}

The MFE beat bifurcation lies at the center of our multi-band oscillation theory and is worth more meticulous investigations.
Here, we present a novel geometrical representation of the bifurcations by {\it projecting} network MFE dynamics on low-dimensional manifolds. 
In our previous study, we developed a first-principle-based coarse-grain model consisting of the total E/I conductances ($g^{E,I}$) and the numbers of subthreshold neurons near the firing threshold $V_{th}$ (E and I ``gate" neurons). 
Though drastically reduced from the high-dimensional spiking network model, this coarse-grain model successfully captured gamma oscillations. In the reduced state space, gamma oscillations are restricted on a nearly 2D manifold except for the initiation of each MFE. 

To describe the geometrical features, we use an analogy of ``root-leaf" structure: ``root" for the MFE initiations and ``leaf" for trajectories of MFEs and post-MFE dynamics represented on the 2D manifold.
Specifically, at the end of each inter-MFE interval, the inertial drives produced by recurrent synaptic couplings mostly dissipate.
Therefore, the network states at each MFE initiation are more governed by the randomness from external stimuli, composing a common area with large stochasticity and dimensionality (``roots"). 
The states of roots were found to dictate the following dynamics of MFEs.  
On the other hand, despite the nonlinear and highly random evolution, MFEs and post-MFE dynamics exhibited sufficiently low dimensional structures that the trajectories are restricted as loops on a 2D manifold (``leaves"). We refer readers to \cite{CaiWuTaoXiao2021} for the rest of the details.

The 2D manifold depicting Gamma oscillations also captures multi-band oscillations by its ``root-leaf" structure. 
Notably, the MFE beats form a one-to-one mapping to the different branches of leaves: When multiband oscillations are illustrated in the same state space, we find that the number of ``leaf" clusters equals to the MFE beat number. Furthermore, the emergence of multiple leaves is dictated by the MFE beat bifurcations (Fig.~\ref{Fig5: Manifold}). 
To visualize different branches of leaves, we dissect the trajectories of network dynamics on $I_n$ (as shown in Sect.~\ref{Sect-3.2.1-MFE_Mapping}), then color each section according to $m_n$ (weak MFEs are specially colored in dark blue).
On the MFE manifold, similarly colored trajectories cluster together on the same leaves, signaling an overall bifurcation structure. 

For each MFE beat bifurcation map in Fig.~\ref{Fig4: Probability density map}, we select four points in the scanned 1D parameter space and demonstrate network dynamics in the reduced space. 
Taking $S^{EI}$ for example (Fig.~\ref{Fig5: Manifold}A), recall the ``1-3-2" pattern of the bifurcation map. Correspondingly, for the 1-beat rhythm, the trajectories of $S^{EI}=0.0196$ only form one leaf with light green ($m \approx 0$, Fig.~\ref{Fig5: Manifold}A first plot). 
In the second and third plots, the three-beat rhythms are represented by the three separate leaves (dark blue, orange, and light green) in the reduced space. 
For $S^{EI}=0.0214$, the 2-beat rhythm is echoed by the merging of the two large leaves.
Similar observations of $S^{\rm ext}$ and $p$ (Fig.~\ref{Fig5: Manifold}BC) confirm the relation between network randomness and ``leaf" branches.
Surprisingly, for different refractory periods, the 2D manifold capturing network dynamics undergoes substantial changes (Fig.~\ref{Fig5: Manifold}D). The reduced space reveals that the single-neuron dynamics, especially the parameters governing physiological timescales, play a crucial role in the emergence of MFE beats and multi-band oscillations.

\subsection{MFE beats captured by reduced ODE models}
\begin{figure*}
  \begin{center}
    \includegraphics[width=\textwidth]{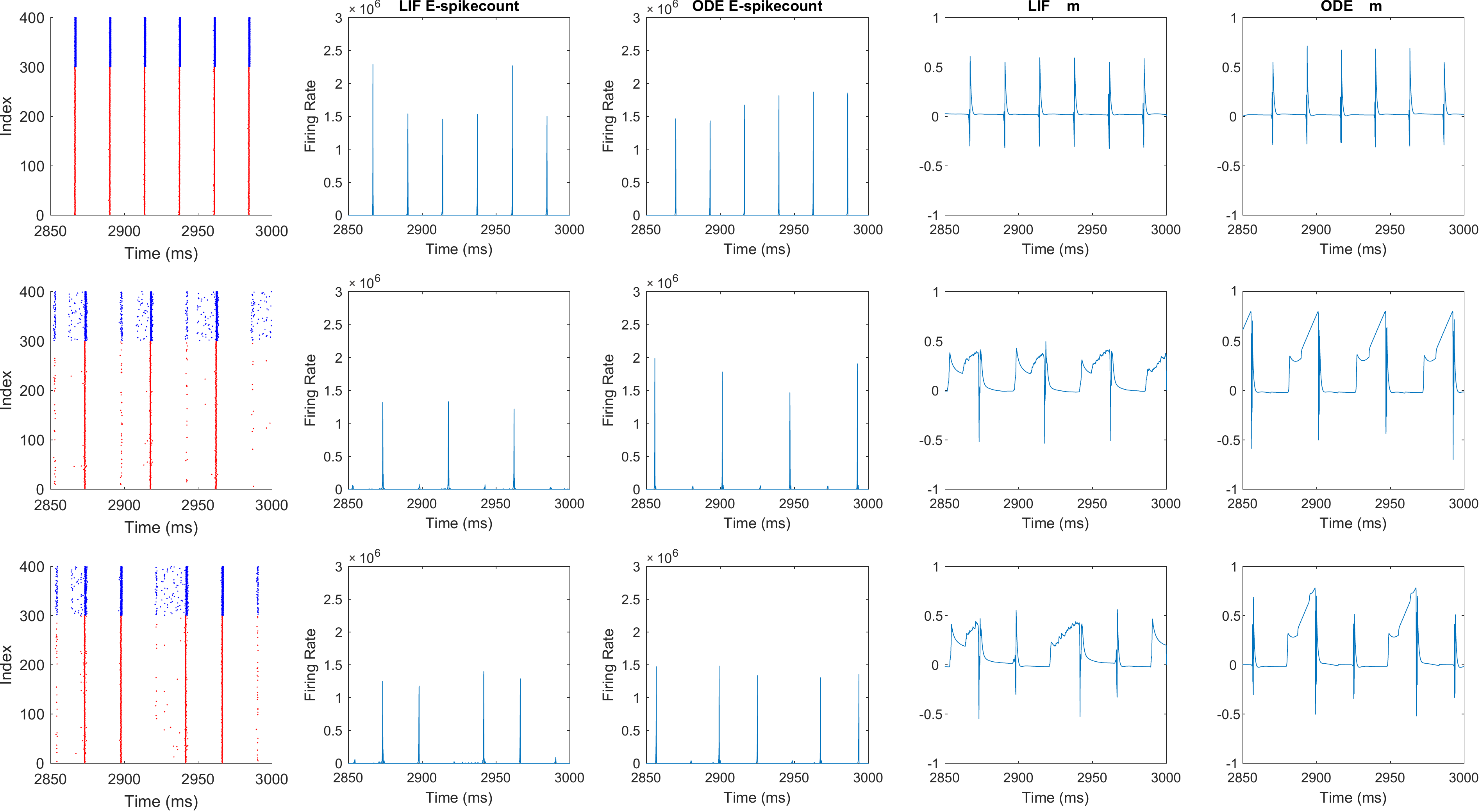}
    \caption{Comparison between the dynamics generated by the spiking neuronal network and the reduced ODE model. Row 1-3 displays the MFE beats produced by both models, where the parameter sets are chosen the same as Fig.~\ref{Fig3: Iters}. \textbf{Column 1}: Raster plots of the spiking neuronal network. \textbf{Column 2\&3}: Firing rates of both models. \textbf{Column 4\&5} Differences between the averages of E/I potential profiles ($m$) of both models.} 
    \label{Fig5: Gamma Features}
  \end{center}
\end{figure*}

\begin{figure}
  \begin{center}
    \includegraphics[width=0.5\textwidth]{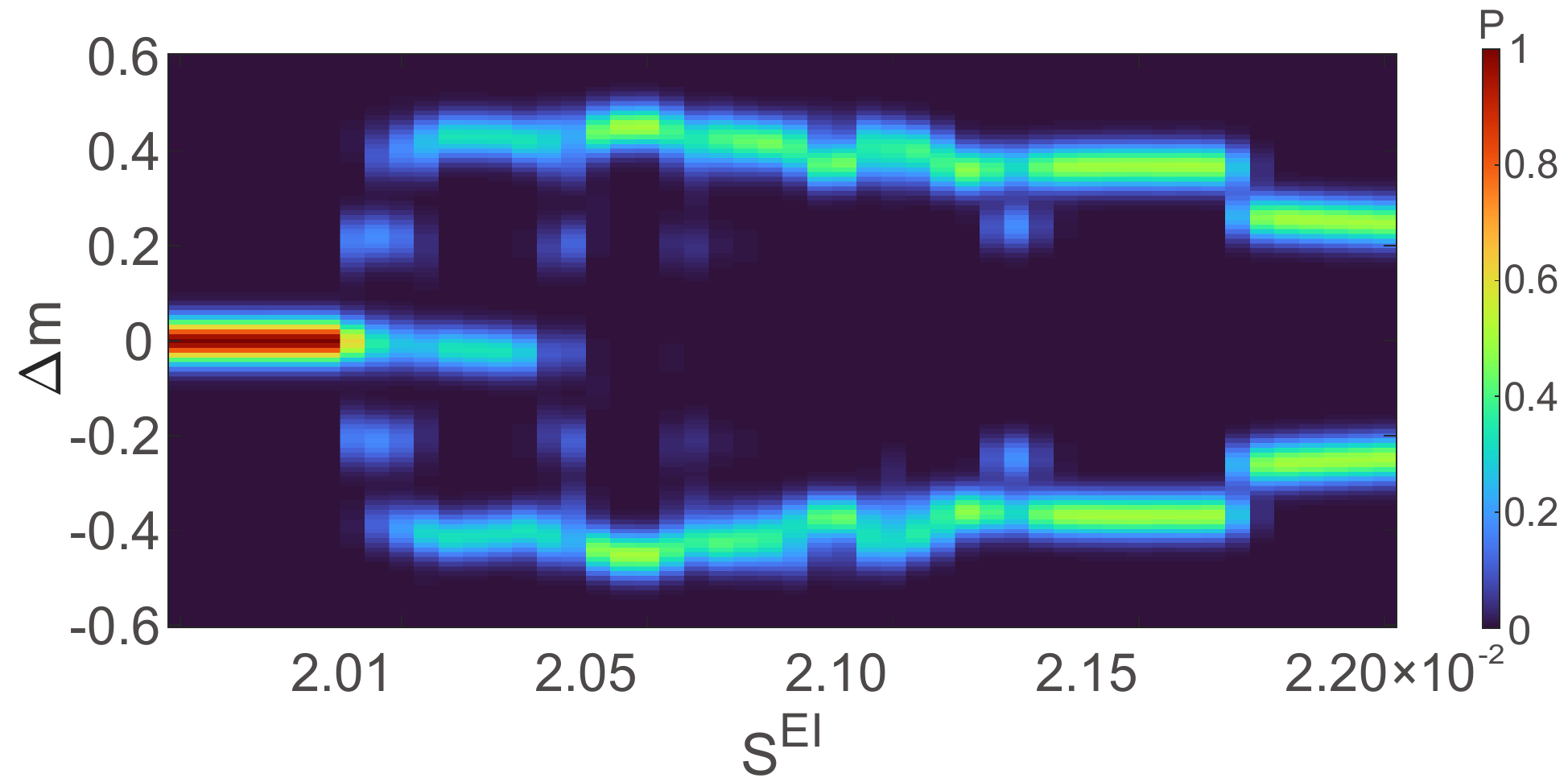}
    \caption{The bifurcation map of MFE beats produced by the ODE model. Only cortical synaptic coupling strength $S^{EI}$ is varied, and this map is comparable to Fig.~\ref{Fig4: Probability density map}A.} 
    \label{Fig7: ODE bifur map}
  \end{center}
\end{figure}
We have been studying the MFE beats by a combination of first principles and data-driven model reductions. 
Although the reduced mapping $f$ provides a solid explanation for the emergence of multi-band oscillatory dynamics, implicitly it also shares the common weaknesses of data-driven models. Specifically, the soundness of the reduced mapping $f$ is dictated by two factors: First, the validity of the model reduction hypothesis; and second, the size and quality of the surrogate data from network simulations. The former leads to a drastic reduction of the complicated E-I competition to a scalar ($m_n$), requiring more careful investigation to make sure it is not a ``blind trust". 
For the latter, the reliability of $f$ is weakened by the large fluctuation in some surrogate data. On the other hand, a large size of high-quality surrogate data from network simulation can be computationally expensive, and even call into question the meaning of the reduced model itself.

To overcome the weakness mentioned above, we develop an \textit{autonomous} ODE model based only on the averages and variances of E/I potential profiles. The ODE model is governed by the mean-field representations of E-I competitions combined with population fluctuations. Opposed to the previous data-driven model, this ODE model is completely based on first principles and free of any surrogate data, hence independent from the numerical simulations of network dynamics. 

The key idea of the ODE model is to approximate the distributions of E/I potential profiles with multiple (instead of one) Gaussian peaks to echo the observation that E/I populations are only partially involved in the weak MFEs. 
During the evolution of the model, the potential profiles will undergo splitting and merging, which are caused by the threshold-crossing during MFE and reactivation after MFE. Specifically, the potential profile of a neural population would be split if a fraction of neurons fire during an MFE but the others remain subthreshold. The potential profiles of the spiked neuron would reset to $V_{r}$ and be modeled as an additional truncated Gaussian in the probability density function. On the other hand, we assume two peaks would merge into one if they get close enough. Each peak is described as a truncated Gaussian due to the same reason as above (both E/I-neurons are primarily driven by external Poissonian inputs in the absence of recurrent stimuli). In all, the autonomous ODE model describe the E-I competition by the dynamics of the splitting and merging of Gaussian peaks. We refer readers to \textbf{Methods} for more details.

Crude as it is, the autonomous ODE model provides a remarkably good approximation to MFE beats, especially when the variance of the each Gaussian peak is small.
We find that ODE model produces similar crucial dynamical features as the MFE beats produced by the network (Fig.~\ref{Fig5: Gamma Features}), including firing rates, firing synchronicity, and oscillation frequency.  
Most importantly, the ODE model also exhibits qualitatively similar dynamics of potential profile averages ($\Delta m$).
We selectively demonstrate the performances of the ODE model for the three parameter sets displayed in Fig.~\ref{Fig3: Iters}.
Furthermore, compared to Fig.~\ref{Fig4: Probability density map}A, the ODE model generates similar bifurcation map of MFE beats when the cortical synaptic coupling weights varies (Fig.~\ref{Fig7: ODE bifur map}).

\begin{figure*}
  \begin{center}
    \includegraphics[width=\textwidth]{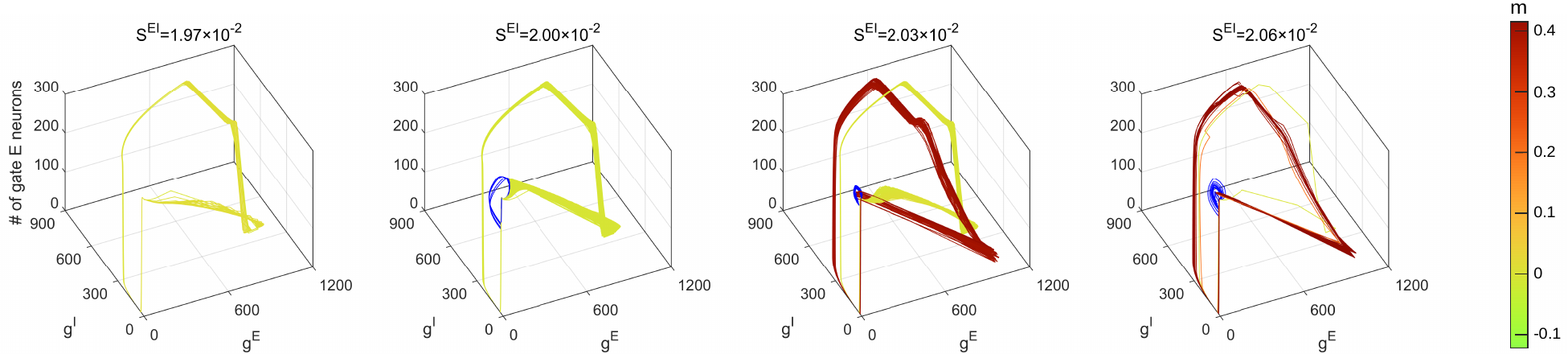}
    \caption{Similar as Fig.~\ref{Fig5: Manifold}A, multi-band oscillations produced by the autonomous ODE model are presented in the same reduced 3D space.} 
    \label{Fig8: Manifold-ODE}
  \end{center}
\end{figure*}

Notably, the ODE model reveals a similar ``root-leaf" structure and leaf branching during MFE bifurcations (Fig.~\ref{Fig8: Manifold-ODE}) when illustrated in the same reduced space concerning E and I ``gate" neurons and conductances. 
Since the ODE model evolves the distributions of E/I potential profiles, one can deduce the numbers of ``gate" neurons from the numbers and locations of the Gaussian peaks.  
Remarkably, not only the number of ``leaves" matches the number of MFE beats, but the ODE model trajectories shares similar geometrical features with the representation of the network simulation (Fig.~\ref{Fig5: Manifold}).
This suggests that the complex network dynamics can be successfully captured by considering only the most important features of potential profiles, even though the splitting and merging of Gaussian peaks in the ODE model are very coarse descriptions of the former.

Taking together, the autonomous ODE model supports our main conclusions that
\begin{enumerate}
    \item Multi-band oscillation, i.e., the various MFE beats, emerge from the E-I competition in different fashions;
    \item The E-I competition in network dynamics can be captured by the E/I potential profiles, and further sufficiently represented by their important features.
\end{enumerate}

\section{Discussion\label{sec_discs}}

The emerging picture of brain cognition is one in which coherent neural activities form the physiological basis of many cognitive functions. Prominent amongst the patterned population activities are the various oscillatory rhythms which have been implicated in sensory, perceptions, attention, and learning and memory. While the experimental investigations of coherent neural activities have formed the theoretical underpinnings of information coding in the brain \cite{SalinasSejnowski2001,ColginEtAl2009,quiroga2009extracting,QuianQuirogaPanzeri2013,sigala2014role,herweg2016theta}, the understanding of how neural circuits can generate patterned activity remain incomplete. Previous theoretical studies have largely focused on single-band oscillatory activities. In addition to the theories for Gamma oscillations referred to earlier in this manuscript
, various theoretical studies the mechanism behind theta \cite{varga2008presence,hangya2009gabaergic,bezaire2016interneuronal,Colgin2016,zhang2018theta,segneri2020theta}, alpha \cite{zhang2018theta,zonca2021emergence}, beta oscillations \cite{pavlides2015computational,tan2016post,yu2019oscillation}, and others.   By combining these ideas with oscillatory inputs, studies have shown that, in general, multi-band oscillations can emerge near Hopf bifurcations of the driven population \cite{jensen2007cross,segneri2020theta}.

Here we take a different tack. Recent work have shown that a wide variety of coherent activity can be generated by homogeneously structured, fluctuation-driven neuronal networks, even under temporally constant drive. In particular, experimental and simulational studies have demonstrated the ubiquity of self-organized, nearly synchronous population spiking activity of variable size and duration. Theoretically, the main obstacle to a full analysis is a clear delineation of the strong stochastic competition between excitatory and inhibitory sub-populations. Already, the mathematical description of the so-called multiple-firing events needed the development of a specific ensemble average, i.e., a `partitioned ensemble average' that treated MFEs differently from nearly homogeneous spiking patterns \cite{ZhangZhouEtAl2014,ZhangRangan2015} (which can be faithfully captured by classical coarse-graining techniques, e.g, Fokker-Planck approaches \cite{shao2020dimensional}). In recent work, we also uncovered a nearly two-dimensional manifold underlying stochastic gamma oscillations by projecting onto a suitable, low-dimensional state space \cite{CaiWuTaoXiao2021}.

Here we focus on a dynamical regime between single-band, stochastic oscillatory rhythms and general heterogeneous patterned activities. During our exploration of the dynamical possibilities of homogeneous networks that generate MFEs, we found that many parameters led to the emergence of nearly periodic patterns consisting of more than one single frequency. Because of the prevalent periodicity, we found it natural to combine classical Poincaré sectional theories with modern data-driven techniques. After a few steps of model reductions, the complex MFE dynamics were captured by the neuronal potential profiles on the time sections of MFE initiations, which were further computed from a series of Monte Carlo simulations of transient network dynamics. The Poincaré sectional theory revealed the crucial role of the delicate E-I competition in MFE bifurcations, leading to multiple frequency bands in network oscillations.

While a data-driven Poincaré return map was instrumental in identifying the bifurcations behind the emergence of multi-band patterns, a more detailed understanding of the roles played by various dynamical variables emerged in our dimensional reduced modeling. Viewed in our framework, a simple geometry of ``leaves and roots" described well the multi-band oscillatory regime. When displayed in a suitably reduced state space, the stochastic neuronal oscillations coherently formed a nearly 2D manifold, on which a collection of ``leaves" (describing each MFEs and the post-MFE dynamics) and ``roots" (indicating the complex initial conditions triggering each MFE) delineate the manifold of multi-band oscillations. Furthermore, based on these observations, and through well-motivated assumptions, we managed to develop an ODE model description that captured the essentials of the population membrane potential dynamics. The multi-band oscillations emerging from spiking networks and the ODE model both exhibits similar geometrical representations of MFE cycles, with the number of MFE beats corresponds to the branches of ``leaves" in the state space.

The modern view of cognition as neuronal population dynamics has also considered these neural network computations as trajectories on manifolds\cite{churchland2010stimulus,ManteEtAl2013,JanakAndTye2015,GallegoEtAl2017,SaxenaAndCunningham2019,CuevaEtAl2020}. Recent dynamical theories of stochastic neuronal network dynamics have progressed beyond f-I curves, static input-output relations, thresholded-linear filters, and weakly nonlinear expansions \cite{BrunelHakim1999,CaiKinetic2006,KeeleyEtAl2019}. Amongst the various neuronal dynamical activities, periodic synchronization has received the most attention, both experimentally and theoretically. We view our work here as complementary to previous studies that focused on identifying mechanisms of synchronization in heterogeneous networks \cite{AcebronEtAl2005,MontbrioEtAl2015,AshwinCoombesNicks2016,BickEtAl2020}. By identifying the population membrane potential distribution as the major player in the generation of multi-frequency oscillations, we have uncovered a novel dynamical route to multi-band oscillations in a simple, homogeneous neuronal network, without appealing to oscillatory inputs or more cell types and synaptic time-scales. The central role of the membrane potential distribution points out the possibility of maintaining and manipulating these multi-frequencied, synchronizing populations, which we hope to investigate more fully in ongoing and future work.

\newpage
\section*{Methods\label{sect_mthd}}

\subsection*{Integrate-and-Fire Network}
We consider an integrate-and-fire (I\&F) neuronal network composed by $N_E$ excitatory neurons ($E$) and $N_I$ inhibitory neurons ($I$). For neuron $i$, its membrane potential ($v_i$) is driven by excitatory and inhibitory (E/I) synaptic currents:
\begin{equation}
\label{sect_mthd:IF}
    \begin{split}
    \ddt{v_i} &= \left(g_i^{\textrm{ext}}+g_i^E\right)\cdot\left(V^E-v_i\right)+g_i^I\left(V^I-v_i\right), \\
    g_i^{\textrm{ext}} &= S_i^{\textrm{ext}}\sum_{\mu_i^{\textrm{ext}}}G^E(t-t_{\mu_i^{\textrm{ext}}}), \\
    g_i^E &= \sum_{\substack{j\in E\\j\neq i}}S_{ij}^{E}\sum_{\mu_j^{E}}G^E(t-t_{\mu_{j}^{E}}), \\
    g_i^I &= \sum_{\substack{j\in I\\j\neq i}}S_{ij}^{I}\sum_{\mu_j^{I}}G^I(t-t_{\mu_{j}^{I}}).
    \end{split}
\end{equation}
In Eq.~\ref{sect_mthd:IF}, $g_i^{\{\textrm{ext},E,I\}}$ indicate the external, recurrent excitatory and inhibitory conductances of neuron $i$, which are induced by external stimulus ($\mu_i^{\textrm{ext}}$) and E/I-spikes produced by other neurons in the network ($\mu_{j}^{\{E,I\}}$) respectively. 
The corresponding strength of synaptic couplings are denoted by $S_i^{\textrm{ext}}$ and $S_{ij}^{\{E,I\}}$. 
The postsynaptic conductances are considered as temporal convolution of the spiking series via a Green's function,
\begin{equation}
\label{sect_mthd:Spike_Eff}
    \begin{split}
        G^E(t) &= \frac{1}{\tau^E}e^{-\nicefrac{t}{\tau^E}}h(t), \\
        G^I(t) &= \frac{1}{\tau^I}e^{-\nicefrac{t}{\tau^I}}h(t),
    \end{split}
\end{equation}
where $h(t)$ is the Heaviside function. The time constants, $\tau^{\{E,I\}}$, model the synaptic timescales of the excitatory and inhibitory synapses (such as AMPA and GABA \cite{gerstner2014neuronal}), respectively. 

Neuron $i$ fires when its membrane potential $v_i$ reaches the firing threshold $V^{th}$, after which neuron $i$ immediately resets the rest potential ($V^r$) and remains in a refractory state for a fixed time of $\tau^{\textrm{ref}}$.  It is conventional in many previous studies to use reduced-dimensional units and choose $V^{th} = 1$ and $V^r = 0$ \cite{koch2004biophysics,CaiKinetic2006}.  Accordingly, $V^E = \nicefrac{14}{3}$ and $V^I = -\nicefrac{2}{3}$ are the excitatory and inhibitory reversal potentials.  

While Eq.~(\ref{sect_mthd:IF}) can model a network with arbitrary connectivity structure, in this paper, we focus on a spatially homogeneous network architecture. That is, whether a certain spike released by a neuron of type $Q$ is received by another neuron of type $Q'$ is only determined by an independent coin flip with a probability $P^{Q'Q}$, where $Q,Q' \in \{E,I\}$. Furthermore, $S_{ij}^{\{E,I\}}$ only depends on the subtypes of the post-/pre-synaptic neuronal pairs, i.e., $S_{ij} = S^{Q'Q}$ if $i$ is a type-$Q'$ and $j$ is a type-$Q$ neuron. We point out that this makes any neurons sharing the same potential profile ($v$) interchangeable in the evolution of transient network dynamics.

The choice of network parameters are summarized below:
\begin{itemize}
    \item Frequencies of external input: $\lambda^E = \lambda^I = 7000$ Hz; Strength of external stimuli: $S_i^{\textrm{ext}} = 0.001$.
    \item Strengths of cortical synaptic couplings: $S^{EE} = S^{EI} = S^{II} = 0.02$, and $S^{IE} = 0.008$;
    \item Probability of spike projections: $P^{EE} = 0.15$, $P^{IE} = P^{EI} = 0.5$, and $P^{II} = 0.4$;
    \item Synaptic timescales: $\tau^I = 4.5$ ms.  $\tau^E =1.4$ ms to reflect the fact that AMPA is faster than GABA. 
\end{itemize}

\subsection*{Power spectrum of neuronal oscillations.} The power spectrum density (PSD) measures of the variance in a signal as a function of frequency.  In this study, the PSD is computed as follows:

A time interval $(0,T)$ is divide into time bins $B_n=[(n-1)\Delta t, n \Delta t], n=1,2,...$, the spike density $\mu_n$ per neuron in $B_n$ is given by $\mu_n = \nicefrac{m_n}{N\Delta t}$ where $m_n$ is the total number of spikes fired in bin $B_n$. Hence, the discrete Fourier transform of $\{\mu_n\}$ on $(0,T)$ is given as:  
\begin{align}
    \hat{\mu}(k) = \frac{1}{\sqrt{T}}\sum_{n=1}^{\nicefrac{T}{\Delta t}}\mu_n\Delta t e^{-k\cdot(2\pi i)\cdot(n\Delta t)}.
\end{align}
Finally, as a function of $k$, PSD is the ``power" concentrated at frequency $k$, i.e., $|\hat{\mu}(k)|^2$.

\subsection*{Detection of MFE}
MFE lacks a rigorous definition from previous studies. On the other hand, it is always visually categorized as highly correlated spiking patterns on the temporal rasterplot. In this study, we design an generic algorithm based on the key idea that each MFE is always triggered by recurrent E-E spikes. 
\begin{enumerate}
    \item When more than two E-E spikes shows up in a 2 ms time interval, we set the end of the interval as the initiation of a MFE. 
    \item After a MFE initiation, when less than two E-E spikes are detected in a 2 ms time interval, we set the begining of the interval as the termination of the MFE.
\end{enumerate}
The two steps above temporally dissect the network dynamics into intervals for MFEs and inter-MFE periods. To finalize the MFE detection, we further combine two MFE intervals into one and discard the inter-MFE-interval between them, if the latter is shorter than 1 ms. 
 
It is known that the lengths of MFE and inter-MFE-intervals are related to the net gain of synaptic current from external stimuli \cite{CaiWuTaoXiao2021}. On the other hand, for all parameter regimes we investigated, the majority of inter-MFE-intervals are longer than 10 ms. Although we do not claim it necessarily coherent with all previous studies, we find this simple algorithm captures the great majority of the MFEs in this study and is consistently reliable during our primary investigations to the network parameter space.
  
\subsection*{MFE iteration plots}
To reconstruct the reduced MFE mapping $m_{1} = f(m_0, \xi)$, we perform the following Monte Carlo type simulations of transient network dynamics.
For a fixed value of $m_0$, we randomly draw membrane potentials of every E/I neuron in the network, from two distributions $D_E\sim\mathcal{N}(m_E,\sigma^2_E)$ and $D_I\sim\mathcal{N}(m_E-m_0,\sigma^2_I)$. The variances of both distributions are collected from all times sections of MFE initiations during a 100s simulation of network dynamics.
These potential profiles are treated as initial conditions in the subsequent simulations of the network dynamics until the detection of the second MFE, at which the E/I distributions of potential profiles are collected. 
After that, we compute the difference between the averages of E/I potential profiles ($\hat{m}_1$) as one output of the MFE mapping $f$.
50 simulations are carried out for each $m_0$, contributing to to the data points in the iteration map on the plane of $(m_0,\hat{m}_1)$. 
The probabilistic output of $f(m_0, \xi)$ is therefore estimated as unimodal or multimodal Gaussians from the empirical distributions of $\hat{m}_1$, producing a data-driven reconstruction of the MFE mapping as $\hat{f}$.

The invariant measures of mapping $\hat{f}$ and $\hat{f}-\mathbbm{1}$ ($\mathbbm{1}$ stands for identity function) are denoted as $\chi_{\hat{f}}$ and $\chi_{\hat{f}-\mathbbm{1}}$, and computed by the average of last 500 iterations from a 1000-time iteration of $\hat{f}$.

\subsection*{ODE model}
From the LIF neuronal network model, our ODE model is a further coarse-graining type reduction. 
First of all, due to the interchangeability between neurons with the same potential profiles, the distributions of E/I neuronal membrane potential profiles ($\rho^{E,I}$) are sufficient for the evolution of network dynamics (rather than tracing $v_i$ for each neuron).
Then, we approximate the distribution of neuronal membrane potential profiles by sum of a collection of truncated Gaussian peaks, and the ODE model is governed the splitting and merging of the peaks. 
According to classical results [CaiTaoetal06], the E/I potential profiles are well approximated by Gaussian functions whose variances are dictated by the noise introduced by the external and recurrent input during the absences of MFEs. Readers should note that the following assumptions of the ODE model are designed to reproduce the MFE dynamics in a qualitative and drastically simplified fashion, instead of mimicking a rigorous evolution of a Fokker-Planck type equation concerning the distributions of potential profiles.

\heading{Splitting.} We first assume $\rho^{E,I}$ are both initially Gaussian which are truncated at 3$\sigma$. 
Being stimulated by external drive, the membrain potentials increases and their distributions $\rho^{E,I}$ may cross the threshold the firing threshold $V^{th}$, hence be divided into two parts.
We further assume that the probability mass passing threshold are released from $V^r$ as a new truncated Gaussian peak after the refractory period (and neurons passing threshold forms the MFE).
On the other hand, the remaining probability mass can be pushed back from the threshold due to the inhibition produced by the I-spikes involved in the MFE. For simplicity, the shape of the distribution of the remaining probability mass is fixed during the suppression, i.e., it remains as a truncated Gaussian peak minus the right tails that already crossed the threshold.  
In all, $\rho^{E}$ and $\rho^{I}$ (which are originally unimodal) are divided into two parts after splitting: a Gaussian peak without tail on the high side and another Gaussian peak released from $V^r$. Splitting may occur for a number of times before the merging of any two peaks (defined momentarily), producing multiple truncated Gaussian peaks (with or without tails on the high side). 

\heading{Merging.} Two Gaussian peaks are merged manually when they share common support. The original two Gaussian peaks are replaced by a new Gaussian peak whose mean and variance are averaged from the two merged peaks.

Therefore, the evolution of the distributions of potential profiles is now simplified to the averages and variances of (multiple) Gaussian peaks. For E-neuron profiles, we use $D^E_1,D^E_2,\dots,D^E_{k^E(t)}$ to denote those normal distributions. Here, $k^E(t)$ stands for the number of peaks contained by the distribution and $D^E_i \sim \mathcal{N}(m_{E,i}, \sigma^2_{E,i})$ for $i=1,2,\dots, k^E(t)$. The peaks are ranked in a descending order of their averages. Note that the first peak $D^E_1$ may lack its right tail due to threshold-crossing during MFEs. The probability mass of each peak are $p^E_1,p^E_2,\dots, p^E_{k^{E}(t)}$. 
Similarly, we use $D^I_1,D^I_2,\dots,D^I_{k^I(t)}$ and $D^I_i\sim \mathcal{N}(m_{I,i}, \sigma^2_{I,i})$ to denote the Gaussian peaks of I-neuron profiles and $p^I_1,p^I_2,\dots, p^I_{k^{I}(t)}$ to denote their probability mass. 

During the absence of the spliting and merging, each peak is independently driven by external and recurrent stimulus. Therefore, the evolution of averages and variances of each peak are dipicted by the following ODEs:
\begin{equation}
\label{sect_mthd:ODE} 
\begin{aligned}
&\left \{
\begin{split}
\ddt{\sigma^2_{E,i}} &= \lambda^E (S^{\textrm{ext}})^2 - 2\cdot \frac{\sigma^2_{E,i}}{V^{th}-V^I} \frac{S^{EI}}{\tau^{I}}g^{EI},\\
       \ddt{m_{E,i}} &= \lambda^E S^{\textrm{ext}} + \frac{S^{EE}}{\tau^{E}}g^{EE} - \frac{S^{EI}}{\tau^{I}}\frac{m_{E,i}-V^I}{V^{th}-V^I} g^{EI},\\ 
\end{split}
\right . & \\
&\left \{
\begin{split}
\ddt{\sigma^2_{I,i}} &= \lambda^I (S^{\textrm{ext}})^2 - 2\cdot \frac{\sigma^2_{I,i}}{V^{th}-V^I} \frac{S^{II}}{\tau^{I}}g^{II},\\
       \ddt{m_{I,i}} &= \lambda^I S^{\textrm{ext}} + \frac{S^{IE}}{\tau^{E}}g^{IE} - \frac{S^{II}}{\tau^{I}} \frac{m_{I,i}-V^I}{V^{th}-V^I} \cdot g^{II},
\end{split}
\right . &\\
\end{aligned}
\end{equation}
In Eq.~\ref{sect_mthd:ODE}, we assume that the variances of each peak are dominated by the noise from external and recurrent inhibition. The term $\frac{m_{Q,i}-V^I}{V^{th}-V^I}$ normalize the contribution of I current driving of $m_{Q,i}$ due to the conductance-based setup in Eq.~\ref{sect_mthd:IF}, where $Q\in\{E,I\}$. The corresponding normalization term for E current is saved for simplicity since $\frac{m_{Q,i}-V^E}{V^{th}-V^E}$ are closer to 1. 


\section*{Acknowledgments} This work was partially supported by the Natural Science Foundation of China through grants 31771147 (T.W., R.Z., Z.W., L.T.) and 91232715 (L.T.). Z.X. is supported by the Courant Institute of Mathematical Sciences through Courant Instructorship.

We thank Lai-Sang Young and David McLaughlin (New York University) and Yao Li (The University of Massachusetts Amherst) for their useful comments.

\section*{AUTHOR DECLARATIONS}
\subsection*{Conflict of Interest}
The authors have no conflicts to disclose.
\subsection*{Data Availability Statement}
Pending data and code available upon the acceptance of the manuscript.
\bibliography{MultibandBib}

\end{document}